\documentclass[prl,twocolumn,superscriptaddress]{revtex4-1}
\usepackage{graphicx}
\usepackage{physics}
\usepackage{bm, amsmath, amssymb, braket}
\usepackage{times}
\usepackage{multirow}
\usepackage{ascmac}
\usepackage{amsthm}
\usepackage{float}
\usepackage{color}
\usepackage{comment}
\usepackage{mathtools}  
\usepackage{mathrsfs} 
\usepackage[colorlinks=True,urlcolor=blue,linkcolor=blue,citecolor=blue]{hyperref}
\usepackage{xcolor}
\usepackage{setspace}
\usepackage{subfigure}
\usepackage[toc,page]{appendix}

\begin{document}
\title{Altermagnetic Kondo Logic via Floquet Symmetry Conversion}
\author{Haojie Shen}
\affiliation{National Laboratory of Solid State Microstructures and Department of Physics, Nanjing University, Nanjing 210093, China}
\author{Xinchen Zhou}
\affiliation{National Laboratory of Solid State Microstructures and Department of Physics, Nanjing University, Nanjing 210093, China}
\author{Baigeng Wang}
\affiliation{National Laboratory of Solid State Microstructures and Department of Physics, Nanjing University, Nanjing 210093, China}
\affiliation{Collaborative Innovation Center of Advanced Microstructures, Nanjing University, Nanjing 210093, China}
\affiliation{Jiangsu Physical Science Research Center, Nanjing University, Nanjing 210093, China}
\author{Rui Wang}
\email{rwang89@nju.edu.cn}
\affiliation{National Laboratory of Solid State Microstructures and Department of Physics, Nanjing University, Nanjing 210093, China}
\affiliation{Collaborative Innovation Center of Advanced Microstructures, Nanjing University, Nanjing 210093, China}
\affiliation{Jiangsu Physical Science Research Center, Nanjing University, Nanjing 210093, China}
\affiliation{Hefei National Laboratory, Hefei 230088, People's Republic of China }

\begin{abstract}
A common belief in Kondo physics is that a magnetic impurity can act as an efficient probe of key properties of its host bath. In altermagnets, however, this intuition fails in a symmetry-enforced way: the momentum-dependent spin splitting cancels in the local impurity spectrum, making a conventional single-impurity Kondo resonance essentially blind to the altermagnetic form factor. Here, we overcome this constraint by introducing Floquet symmetry conversion. We show that an in-plane AC field converts the hidden altermagnetic spin splitting into a locally detectable spin-dependent hybridization channel, which is further amplified by Kondo correlations into a novel altermagnetic Kondo effect. Remarkably, the Kondo splitting exhibits a nontrivial dependence on the field orientation and inherits the same crystalline form factor as the altermagnetic host, thereby serving as Kondo tomography of altermagnetic spin splitting. This field-controlled response further provides a natural route to symmetry-controlled Kondo logic gates. Our results establish a field-controllable altermagnetic Kondo effect, linking symmetry-resolved impurity spectroscopy to Kondo logic functionality.

\end{abstract}

\maketitle
\emph{\color{blue}{Introduction.--}}
Altermagnets constitute a new class of compensated collinear magnets, characterized by momentum-dependent spin splitting in reciprocal space and vanishing net magnetization in real space
\cite{Litvin1974SpinGroups,Hayami2019Momentum,Yuan2020Giant,Wu2007FermiLiquid,Liu2022SpinGroup,Smejkal2022Beyond,Smejkal2022Landscape,Ahn2019RuO2,Hayami2020BottomUp,Yuan2021Prediction}.
Unlike conventional ferromagnets or antiferromagnets, their spin splitting is dictated by spin-group symmetry and transforms as distinct crystalline form factors
\cite{Smejkal2022Beyond,McClarty2024Landau,Fernandes2024Zeeman,
Bhowal2024Octupoles,Chen2024SpinSpaceGroups,Hu2025Catalog,Huang2025SpinInversion}.
This symmetry-enriched spin structure has led to unconventional responses, including anomalous Hall effects
\cite{Smejkal2020Hall,Feng2022Hall,GonzalezBetancourt2023AHE,Reichlova2024AHE},
unusual Josephson effects
\cite{Ouassou2023Josephson,Beenakker2023Andreev},
spin-charge conversion
\cite{Naka2019SpinCurrent,GonzalezHernandez2021SpinSplitter,
Bai2022SpinSplittingTorque,Bose2022TiltedSpinCurrent,
Karube2022SpinSplitterTorque,Bai2023SpinCharge},
and related transport and optical phenomena
\cite{Smejkal2022Magnetoresistance,Rao2024Optical,Liu2026SpinPrecession},
substantially broadening the notion of magnetic order in solids.

Despite this progress, two issues remain central.
First, direct and symmetry-resolved detection of altermagnetic spin splitting remains highly desirable.
Although signatures of altermagnetic band splitting have been reported by angle-resolved photoemission spectroscopy
\cite{Krempasky2024Kramers,Fedchenko2024RuO2,Lee2024MnTe,
Osumi2024MnTe,Reimers2024CrSb,Ding2024CrSb,Yang2025Mapping,
Jiang2025Metallic,Zhang2025SpinValley}
and spin-resolved scanning tunneling microscopy
\cite{Wang2025AtomicSpinSensing},
resolving the underlying symmetry form factor remains challenging.
Second, most studies focus on single-particle properties, while correlation effects built on altermagnetic band structures remain much less explored
\cite{Diniz2024Kondo,Sato2024CorrelationHall}.
Although not intrinsic to all altermagnets, strong correlations can become important in strongly coupled bands
\cite{Daghofer2026Polarons},
heavy-fermion systems
\cite{Zhao2025Altermagnetism},
altermagnet-based heterostructures
\cite{Zhu2026Proximity},
and doped altermagnets
\cite{Mazin2021FeSb2}.
It is therefore natural to ask whether altermagnetic spin splitting can appear in correlated many-body phenomena and whether such phenomena can reveal its intrinsic symmetry.

\begin{figure}[t]
    \centering
    \includegraphics[width=\columnwidth]{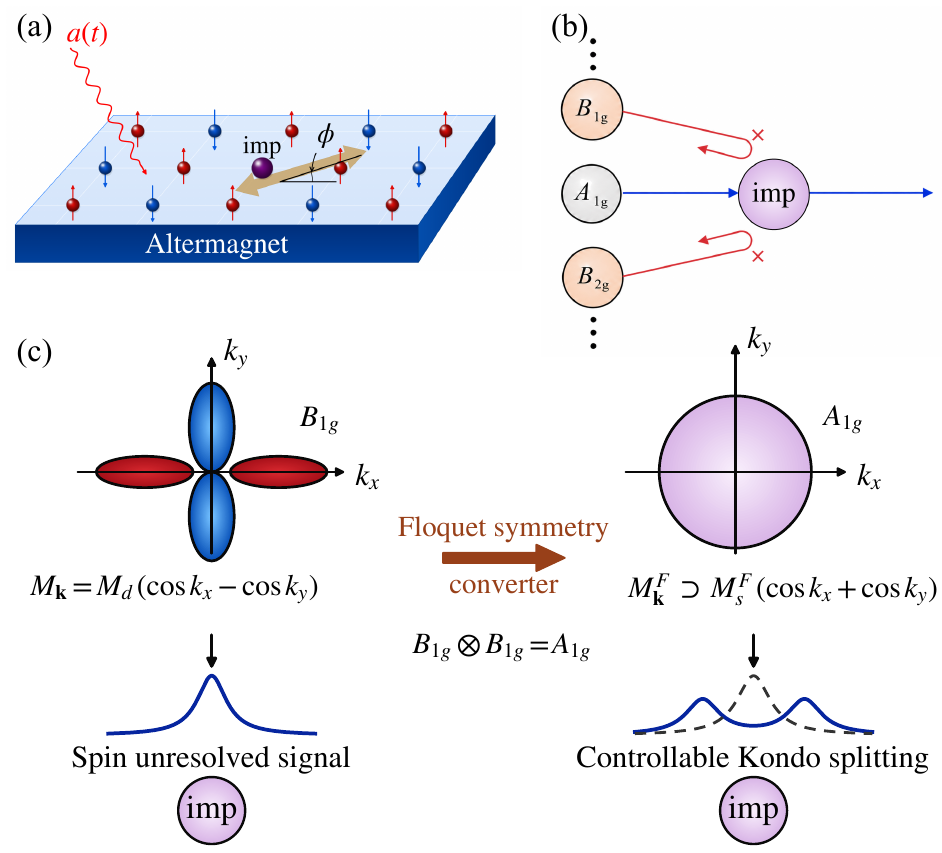}
    \caption{Floquet symmetry conversion. (a) An impurity couples to an altermagnet driven by an in-plane AC field at orientation $\phi$. (b) An $s$-wave impurity projects the bath onto the $A_{1g}$ channel and is therefore blind to nontrivial form factors such as $B_{1g}$ and $B_{2g}$. (c) For a $B_{1g}$ $d$-wave altermagnet, the matching $B_{1g}$ drive tensor generates an $A_{1g}$ component detectable as a field-controlled Kondo splitting.}
    \label{fig:overview}
\end{figure}

Kondo impurities offer a particularly sharp probe of this question
\cite{Kondo1964,Wilson1975},
but also expose a symmetry constraint.
A local impurity with isotropic hybridization couples only to the local
$A_{1g}$ component of the host electrons \cite{footnote0}.
Consequently, nontrivial altermagnetic form factors, including the
$B_{1g}$ and $B_{2g}$ components of $d$-wave altermagnets, cancel in the
impurity hybridization and remain invisible to a single $s$-wave impurity
\cite{Diniz2024Kondo,Lee2023Impurities}
[Fig.~\ref{fig:overview}(b)].
Its Kondo resonance is therefore expected to be essentially
indistinguishable from that in a normal metal.
Recent quasiparticle-interference studies
\cite{Hu2025QPI,Petermann2025SpinQPI}
and two-impurity Kondo proposals \cite{Qin2026TwoImpurity}
provide valuable probes of the altermagnetic form factor, but rely on
spatial interference patterns whose interpretation may be affected by
surface disorder, Fermi-surface geometry, and lifetime broadening.
More fundamentally, this raises a question central to Kondo physics:
can a strictly local Kondo impurity access crystalline-symmetry
information encoded nonlocally in momentum space?
An affirmative answer would extend the conventional role of a Kondo
impurity from probing local bath properties to resolving the symmetry
structure of a momentum-dependent magnetic state.

In this Letter, we reveal a novel altermagnetic Kondo phenomenon in a single-impurity response, enabled by Floquet symmetry conversion.
We show that an in-plane AC electric field reshapes the altermagnetic form factor and generates a locally visible $A_{1g}$ spin-dependent hybridization channel
[Fig.~\ref{fig:overview}(c)].
The spin splitting hidden from a static local impurity is thereby converted into a finite local spin contrast, which Kondo correlations amplify into a spin-resolved many-body resonance.
The corresponding Kondo peak becomes controllable and can be tuned between unsplit and split regimes [Fig.~\ref{fig:overview}(c)].
Unlike conventional Kondo splitting induced by a magnetic field
\cite{Zhang2010KondoSplitting}
or a spin-polarized perturbation
\cite{Niu2015KondoSplitting,Qi2008Kondo},
the present splitting originates from the symmetry-converted altermagnetic bath and exhibits an orientation-dependent spin contrast that directly inherits the underlying form factor.
The single-impurity Kondo resonance therefore serves as a many-body tomography of altermagnetic spin splitting, resolving not only its presence but also its symmetry origin.

This field-controlled response further provides a natural route to logic functionality.
Assigning the split and unsplit resonances as two output states allows the altermagnetic symmetry to determine the truth table under different field configurations. This leads to a symmetry-selected binary response that we refer to as altermagnetic Kondo logic.
In representative $d$-wave altermagnets, $B_{1g}$ and $B_{2g}$ form factors realize XOR and AND operations, respectively, promoting the impurity resonance from a passive spectroscopic probe to an active symmetry-controlled logic element.
The uncovered Kondo phenomenon therefore not only diagnoses altermagnetic symmetry but also opens a device-oriented route toward logic functionality, advancing fundamental knowledge of Kondo physics.

\textit{\color{blue}{Floquet symmetry converter.--}} We consider a metallic altermagnet  with well-defined  spin-resolved Fermi surfaces. Its low-energy excitations are described by \cite{Smejkal2022Beyond}, $H_0(\mathbf{k})=\sum_{\mathbf{k}\sigma}(\xi_{\mathbf{k}}+M_{\mathbf{k}}\sigma)c^{\dagger}_{\mathbf{k}\sigma}c_{\mathbf{k}\sigma}$, where $\xi_{\mathbf{k}}$ is the spin-independent dispersion, $M_{\mathbf{k}}$ describes the momentum-dependent altermagnetic spin splitting, and $\sigma=\uparrow,\downarrow$ denotes spin up and down. Multiband and multiorbital effects are not essential to the symmetry mechanism discussed below and will be omitted for clarity. The altermagnetic spin texture is encoded in $M_{\mathbf{k}}$, which transforms according to an irreducible representation (irrep) $\Gamma_{\mathrm{AM}}$ of a point group $G$:
\begin{equation}\label{irrep}
    M_{g^{-1}\mathbf{k}}=\chi_{\Gamma_{\mathrm{AM}}}(g)M_{\mathbf{k}},\qquad \forall g\in G.
\end{equation}
Here we focus on the one-dimensional (1D) real irreps relevant to altermagnets, and $\chi_{\Gamma_{\mathrm{AM}}}(g)$ is the corresponding character. For instance, the spin splitting of a $d$-wave altermagnet in a tetragonal crystal may transform as $B_{1g}$ or $B_{2g}$ of $D_{4h}$. Since the spin-resolved structure of an altermagnet is encoded in $M_{\mathbf{k}}$, a local and symmetry-sensitive probe of this form factor is highly desirable.  A possible way to achieve this is to examine the impurity physics in altermagnets.

We first show that an ideal on-site impurity is symmetry-blind to such a nontrivial altermagnetic form factor. Consider an Anderson impurity coupled locally to the AM bath through an isotropic $s$-wave hybridization $V_0$, $H_{\mathrm{hyb}}=V_0\sum_{\sigma}\int\frac{d\mathbf{k}}{(2\pi)^2}(c^{\dagger}_{\mathbf{k}\sigma}d_{\sigma}+d^{\dagger}_{\sigma}c_{\mathbf{k}\sigma})$. The bath enters the impurity problem through $\Delta_{\sigma}(z)=\frac{|V_0|^2}{4\pi^2}\int_{\mathrm{BZ}}d\mathbf{k}(z-\xi_{\mathbf{k}}-\sigma M_{\mathbf{k}})^{-1}$, where $z$ is a complex frequency. Its spin-dependent part is $\delta\Delta(z)=\Delta_{\uparrow}(z)-\Delta_{\downarrow}(z)=\frac{|V_0|^2}{2\pi^2}\int_{\mathrm{BZ}}d\mathbf{k}\frac{M_{\mathbf{k}}}{(z-\xi_{\mathbf{k}})^2-M^2_{\mathbf{k}}}$. Since the denominator is fully symmetric, this can be written as \cite{sup} $\delta\Delta(z)=\int_{\mathrm{BZ}}d\mathbf{k}M_{\mathbf{k}}W(\mathbf{k})$, with an $A_{1g}$-symmetric weight satisfying $W(g\mathbf{k})=W(\mathbf{k})$. If $M_{\mathbf{k}}$ belongs to a nontrivial irrep, a symmetry operation with $\chi_{\Gamma_{\mathrm{AM}}}(g)\neq1$ makes the Brillouin-zone integral vanish, yielding $\delta\Delta(z)=0$. A local $s$-wave impurity is therefore blind to the altermagnetic spin features even though the bulk bands are spin split \cite{Smejkal2022Beyond,Diniz2024Kondo}. Equivalently, the impurity projects the bath onto the $A_{1g}$ channel, while $d$-wave $B_{1g}$, $B_{2g}$, and $g$-wave $A_{2g}$ form factors have no local $A_{1g}$ projection [Fig.~\ref{fig:overview}(b)].

We now introduce Floquet symmetry conversion as a way to overcome this local symmetry obstruction. An in-plane off-resonant AC electric field is represented by $\mathbf{E}(t)=-\partial_t\mathbf{a}(t)$ with $\mathbf{a}(t)=a_0\cos\Omega t(\cos\phi,\sin\phi)$, where $\phi$ is the field orientation relative to the crystalline $x$ axis [Fig.~\ref{fig:overview}(a)], $\Omega$ the driving frequency, and $a_0$ the field strength. At leading order in the high-frequency expansion, the Peierls shift dresses the altermagnetic form factor into its cycle average $\langle M_{\mathbf{k}+\mathbf{a}(t)}\rangle_T$ \cite{Oka2009Floquet,Bukov2015Floquet,Eckardt2017Floquet,footnote1}. The spin-dependent hybridization function becomes
\begin{equation}\label{drive_split}
    \delta\tilde{\Delta}(z,\mathbf{a})=\int_{\mathrm{BZ}}d\mathbf{k}\,\langle M_{\mathbf{k}+\mathbf{a}(t)}\rangle_T\tilde{W}_{\mathbf{a}}(\mathbf{k}),
\end{equation}
where $\tilde{W}_{\mathbf{a}}(\mathbf{k})$ is a Floquet-renormalized weight \cite{sup}, $\langle\cdots\rangle_T$ denotes the period average, and $\mathbf{a}=(a_0,\phi)$ denotes the drive parameters.

Under a simultaneous point-group operation on momentum and drive direction, $\langle M_{g\mathbf{k}+g\mathbf{a}(t)}\rangle_T=\chi_{\Gamma_{\mathrm{AM}}}(g)\langle M_{\mathbf{k}+\mathbf{a}(t)}\rangle_T$, and hence
\begin{equation}\label{constraint}
\delta\tilde{\Delta}(z,g\mathbf{a})=\chi_{\Gamma_{\mathrm{AM}}}(g)\delta\tilde{\Delta}(z,\mathbf{a}).
\end{equation}
Equation~\eqref{constraint} is the central symmetry relation of the Floquet readout. If a drive is invariant under an operation $g$ with $g\mathbf{a}=\mathbf{a}$ and $\chi_{\Gamma_{\mathrm{AM}}}(g)\neq1$, the local spin signal must vanish. For a generic drive direction, however, this constraint is lifted and a finite spin-dependent hybridization is allowed.

The symmetry content of this conversion can be seen by expanding the Peierls-shifted form factor, $\langle M_{\mathbf{k}+\mathbf{a}(t)}\rangle_T=M_{\mathbf{k}}+Q^{(2)}_{i_1i_2}\partial_{k_{i_1}}\partial_{k_{i_2}}M_{\mathbf{k}}/2+\cdots+Q^{(n)}_{i_1\cdots i_n}\partial_{k_{i_1}}\cdots\partial_{k_{i_n}}M_{\mathbf{k}}/n!$, where $Q^{(n)}_{i_1\cdots i_n}=\langle a_{i_1}(t)\cdots a_{i_n}(t)\rangle_T$ is the $n$th-order Floquet drive tensor. This tensor can be decomposed into point-group irreps $\Gamma_{\mathrm{Drive}}$. In $D_{4h}$, the in-plane vector potential transforms as $E_u$, so $\mathrm{Sym}^2(E_u)=A_{1g}\oplus B_{1g}\oplus B_{2g}$ and $\mathrm{Sym}^4(E_u)=2A_{1g}\oplus A_{2g}\oplus B_{1g}\oplus B_{2g}$. Acting on $M_{\mathbf{k}}$, the derivatives generate form factors in $\Gamma_{\mathrm{Drive}}\otimes\Gamma_{\mathrm{AM}}$. A local $A_{1g}$ response therefore appears whenever
\begin{equation}\label{conversion}
    \Gamma_{\mathrm{AM}}\otimes\Gamma_{\mathrm{Drive}}\supset A_{1g}.
\end{equation}
For 1D real representations, this requires $\Gamma_{\mathrm{Drive}}=\Gamma_{\mathrm{AM}}$. Accordingly, $B_{1g}$ and $B_{2g}$ altermagnets are activated by the corresponding components of the quadratic drive tensor, whereas an $A_{2g}$ $g$-wave altermagnet requires a fourth-order tensor. In this sense, the Floquet field acts as a symmetry converter: it supplies the missing irreducible drive tensor and converts a locally canceling altermagnetic form factor into an $A_{1g}$ response detectable by an $s$-wave impurity [Fig.~\ref{fig:overview}(c)]. This mechanism is fundamentally distinct from recent Floquet approaches that engineer bulk spin-splitting textures in antiferromagnets \cite{LiFloquetSpin2026,HuangOddParity2026,ZhuOddParity2026}: here, the drive instead renders a pre-existing altermagnetic form factor locally visible to a quantum impurity.

\textit{\color{blue}{Application to $d$-wave altermagnets.--}} We now specialize the Floquet symmetry conversion mechanism to $d$-wave altermagnets. For a $B_{1g}$ altermagnet with a $d_{x^2-y^2}$ form factor, choosing $g$ as the diagonal mirror operation and the $C_4$ rotation gives, from Eq.~\eqref{constraint}, $\delta\tilde{\Delta}(\pi/2-\phi)=-\delta\tilde{\Delta}(\phi)$ and $\delta\tilde{\Delta}(\phi+\pi/2)=-\delta\tilde{\Delta}(\phi)$. Thus $\delta\tilde{\Delta}(\pi/4)=0$: a linearly polarized AC field along $\phi=\pi/4$ (mod $\pi/2$) leaves the local spin splitting symmetry forbidden. For a generic field orientation, $\delta\tilde{\Delta}(\phi)$ becomes finite and changes sign under a $\pi/2$ rotation. Analogous relations follow for a $B_{2g}$ altermagnet with a $d_{xy}$ form factor, with the nodal directions rotated by $\pi/4$.

For a general $d$-wave altermagnet with mixed $B_{1g}$ and $B_{2g}$ components, the low-energy continuum form factor is $M_{\mathbf{k}}\sim(k_x^2-k_y^2)\cos(2\theta_d)+2k_xk_y\sin(2\theta_d)$, where $\theta_d$ specifies the mixing ratio. For a linearly polarized drive $\mathbf{a}(t)=a_0\cos\Omega t(\cos\phi,\sin\phi)$, its cycle average is $\langle M_{\mathbf{k}+\mathbf{a}(t)}\rangle_T=(a_0^2/2)\cos(2\phi-2\theta_d)$, which is independent of $\mathbf{k}$ and therefore belongs to the local $A_{1g}$ channel. Inserting this Floquet-induced component into Eq.~\eqref{drive_split} yields, in the weak-field continuum limit, $\delta\tilde{\Delta}(z,\mathbf{a})\propto a_0^2\cos(2\phi-2\theta_d)$.

The same mechanism can be demonstrated in a tight-binding lattice model. Consider a $B_{1g}$ $d$-wave altermagnet ($\theta_d=0$), $H_0=\sum_{\mathbf{k}\sigma}\varepsilon^0_{\mathbf{k}\sigma}c^{\dagger}_{\mathbf{k}\sigma}c_{\mathbf{k}\sigma}$, with $\varepsilon^0_{\mathbf{k}\sigma}=-2t(\cos k_x+\cos k_y)-\mu+\sigma M(\cos k_x-\cos k_y)$, where $M$ controls the altermagnetic spin splitting. Under the off-resonant AC field, the leading high-frequency Floquet Hamiltonian is \cite{sup} $H^F=\sum_{\mathbf{k}\sigma}\varepsilon^F_{\mathbf{k}\sigma}c^{\dagger}_{\mathbf{k}\sigma,0}c_{\mathbf{k}\sigma,0}$, where $\varepsilon^F_{\mathbf{k}\sigma}=-2t(\mathcal{J}_x\cos k_x+\mathcal{J}_y\cos k_y)-\mu+\sigma M^F_{\mathbf{k}}$ and $c_{\mathbf{k}\sigma,0}$ denotes the zero-photon sector. The Floquet-renormalized form factor is
\begin{equation}\label{driven_M}
    M^F_{\mathbf{k}}=M^F_d(\cos k_x-\cos k_y)+M^F_s(\cos k_x+\cos k_y),
\end{equation}
where $M^F_d=M(\mathcal{J}_x+\mathcal{J}_y)/2$, $M^F_s=M(\mathcal{J}_x-\mathcal{J}_y)/2$, and $\mathcal{J}_x=J_0(a_0\cos\phi)$ and $\mathcal{J}_y=J_0(a_0\sin\phi)$, with $J_0(x)$ being the zeroth order Bessel function. The first term retains the original $B_{1g}$ symmetry, whereas the second is an $A_{1g}$-symmetric spin-dependent component tuned by $a_0$ and $\phi$. Thus the AC field converts the $B_{1g}$ band structure into a Floquet-dressed one containing a finite local $A_{1g}$ component.

\begin{figure}[t]
    \centering
    \includegraphics[width=\columnwidth]{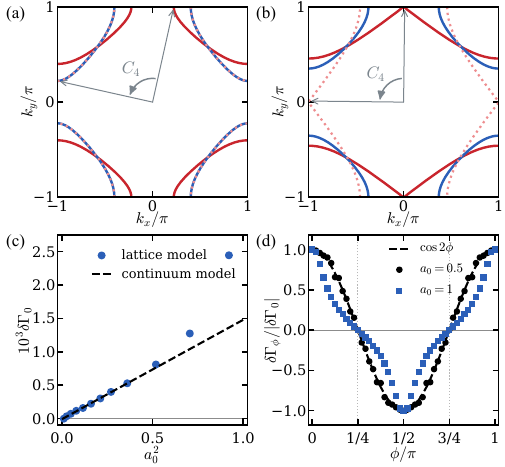}
    \caption{Floquet engineering of the band structure and impurity broadening. (a),(b) Spin-resolved Fermi surfaces without and with a generic drive. (c),(d) $\delta\Gamma_0$ versus the field amplitude and orientation. Parameters are $t=1$, $\mu=1$, $M=0.3$, and $V_0=0.1$. $a_0=0$ for (a), $a_0=1$, $\phi=0$ for (b), and $\phi=0$  for (c).}
    \label{fig:floquet}
\end{figure}

The resulting spin-dependent Fermi surfaces are shown in Figs.~\ref{fig:floquet}(a) and \ref{fig:floquet}(b). Unlike the undriven altermagnet, a $C_4$ rotation no longer maps the spin-up and spin-down Fermi surfaces onto one another for $\phi\neq\pi/4$ (mod $\pi/2$), consistent with the generated $A_{1g}$ component in Eq.~\eqref{driven_M}. We further compute the zero-frequency spin-splitting broadening $\delta\Gamma_0\equiv\delta\Gamma(0,a_0,\phi)$, with $\delta\Gamma(\omega,a_0,\phi)=-\mathrm{Im}\,\delta\tilde{\Delta}(\omega+i0^+,\mathbf{a})$, which measures the difference between the two spin hybridization channels and controls the subsequent Kondo spin asymmetry.

As shown in Fig.~\ref{fig:floquet}(c), $\delta\Gamma_0$ vanishes without driving but becomes finite for $\phi\neq\pi/4$ (mod $\pi/2$). At weak drive it grows linearly with $a_0^2$, while its angular dependence follows the expected $\cos2\phi$ behavior [Fig.~\ref{fig:floquet}(d)]. At larger amplitudes, the full Bessel dressing produces deviations from a pure cosine; nevertheless, the symmetry-enforced nodes at $\pi/4$ (mod $\pi/2$) and sign reversal under a $\pi/2$ rotation remain intact. These nonperturbative features are protected by Eq.~\eqref{constraint}.

\textit{\color{blue}{Kondo tomography of altermagnetic spin splitting.--}} We next show that the Floquet-converted local spin splitting is amplified into a many-body Kondo response. Within the leading high-frequency description, the impurity couples to the zero-photon sector through $H_{\mathrm{hyb}}=V_0\sum_{\sigma}\int\frac{d\mathbf{k}}{(2\pi)^2}(c^{\dagger}_{\mathbf{k}\sigma,0}d_{\sigma}+\mathrm{h.c.})$ \cite{sup}, while $H_{\mathrm{imp}}=\sum_{\sigma}\varepsilon_dd^{\dagger}_{\sigma}d_{\sigma}+Un_{\uparrow}n_{\downarrow}$. The interaction $U$ generates the Kondo correlations.

We solve the resulting Anderson model using full-density-matrix numerical renormalization group (FDM-NRG) \cite{Weichselbaum2007FDMNRG,Bulla2008NRG} and compute the spin-resolved impurity spectra $\mathcal{A}_{\sigma}(\omega)$. The spin-averaged spectrum is $\bar{\mathcal{A}}(\omega)=[\mathcal{A}_{\uparrow}(\omega)+\mathcal{A}_{\downarrow}(\omega)]/2$. We quantify the spectral spin contrast by $P_{\mathcal{A}}(\omega)=[\mathcal{A}_{\uparrow}(\omega)-\mathcal{A}_{\downarrow}(\omega)]/[\mathcal{A}_{\uparrow}(\omega)+\mathcal{A}_{\downarrow}(\omega)]$ and $P_K^+=\omega_c^{-1}\int_0^{\omega_c}d\omega\,P_{\mathcal{A}}(\omega)$, with $\omega_c$ of the order of Kondo temperature, $T_K$. Numerical details and benchmarks are provided in Sec.~II of Supplemental Material.

\begin{figure}[t]
    \centering
    \includegraphics[width=\columnwidth]{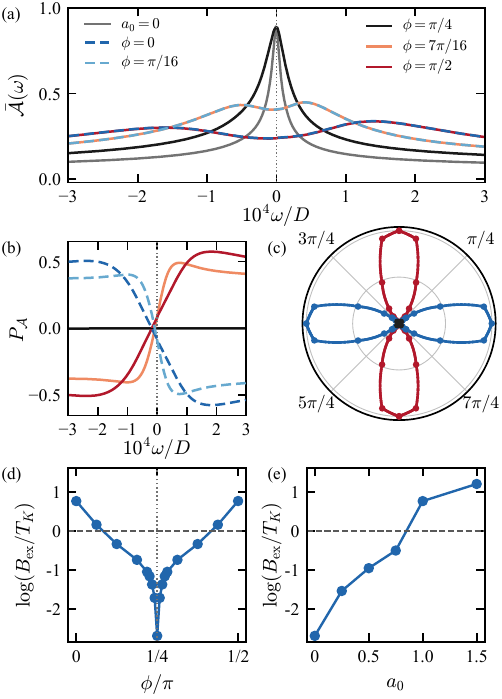}
    \caption{Altermagnetic Kondo behavior for the $B_{1g}$ $d$-wave case. (a) Spin-averaged impurity spectra for different $\phi$. (b) Spectral spin contrast $P_{\mathcal{A}}(\omega)$. (c) Polar plot of $P_K^+$, which inherits the $B_{1g}$ form factor. (d), (e) Effective field $B_{\mathrm{ex}}$ versus orientation and amplitude. Bath parameters are the same as in Fig.~\ref{fig:floquet}; $U/D=0.015$, $\epsilon_d/D=-0.010$. Panels (a--d) use $a_0=1$ except for the $a_0=0$ reference in (a), and (e) uses $\phi=0$.}
    \label{fig:kondo}
\end{figure}

For the $B_{1g}$ tight-binding altermagnet, the undriven case exhibits a sharp Kondo resonance but no spin splitting despite the spin-split bulk bands [Fig.~\ref{fig:kondo}(a)]. The same remains true for the symmetry-forbidden drive direction $\phi=\pi/4$ (mod $\pi/2$), where Eq.~\eqref{constraint} enforces a vanishing local spin signal. At fixed $a_0$, tuning $\phi$ away from this node activates the spin-resolved Kondo response: $\bar{\mathcal{A}}(\omega)$ broadens and eventually evolves into a resolved two-peak structure, while $P_{\mathcal{A}}(\omega)$ becomes increasingly pronounced [Fig.~\ref{fig:kondo}(b)].

The spin contrast also obeys the symmetry constraints inherited from the altermagnetic bath. In particular, $P_{\mathcal{A}}(\omega)$ changes sign between field orientations related by a $\pi/2$ rotation, as illustrated by the $\phi=0$ and $\phi=\pi/2$ curves in Fig.~\ref{fig:kondo}(b). Correspondingly, $P_K^+$ exhibits a $d_{x^2-y^2}$ angular pattern [Fig.~\ref{fig:kondo}(c)]. The Kondo resonance therefore provides a local tomography of the $B_{1g}$ altermagnetic form factor.

\begin{figure}[t]
    \centering
    \includegraphics[width=\columnwidth]{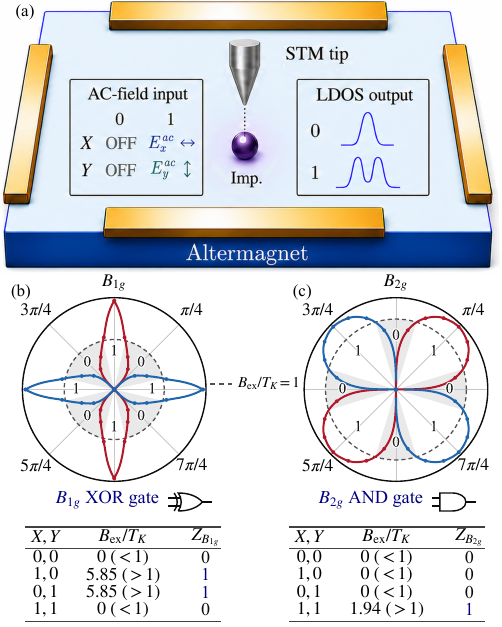}
    \caption{Kondo logic gates. (a) AC fields along $x$ and $y$ serve as inputs, while the impurity LDOS splitting is the output. (b),(c) Split/unsplit regions and truth tables for $B_{1g}$ and $B_{2g}$ altermagnets, realizing XOR and AND gates, respectively.}
    \label{fig:logic}
\end{figure}

A finite spin-dependent hybridization does not immediately imply a resolved Kondo splitting. Although $\delta\Gamma_0$ is nonzero away from the nodal directions [Fig.~\ref{fig:floquet}(c),(d)], the spin-averaged spectrum remains essentially unsplit until the Floquet-induced exchange field competes with the Kondo scale. We estimate this field as $B_{\mathrm{ex}}(a_0,\phi)=\frac{1}{4\omega_c}|\int_{-\omega_c}^{\omega_c}d\omega\,\mathrm{Re}\,\delta\tilde{\Delta}(\omega+i0^+,a_0,\phi)|$. For $B_{\mathrm{ex}}<T_K$, the averaged resonance remains unsplit even though $P_{\mathcal{A}}(\omega)$ detects a finite spin asymmetry; a resolved splitting appears only when $B_{\mathrm{ex}}\gtrsim T_K$. Figures~\ref{fig:kondo}(d) and \ref{fig:kondo}(e) thus demonstrate a controllable crossover between unsplit and split regimes by tuning either $a_0$ or $\phi$.

\textit{\color{blue}{Altermagnetic Kondo logic.--}} The field-controlled Kondo response provides a direct route to symmetry-selected Kondo logic gates. As sketched in Fig.~\ref{fig:logic}(a), two sets of parallel electrodes generate AC electric fields along the crystalline $x$ and $y$ directions, $\mathbf{a}(t)=a_0\cos\Omega t(X\hat{x}+Y\hat{y})$, where $X,Y=0,1$ denote whether the corresponding field component is off or on. The impurity spectrum then exhibits input-dependent Kondo splitting. We assign an unsplit resonance to $Z=0$ and a resolved split resonance to $Z=1$, with the spectral output detectable by STM/STS.

The split or unsplit behavior is controlled by $B_{\mathrm{ex}}(a_0,\phi)/T_K$. For fixed $a_0$, Fig.~\ref{fig:logic}(b)(c) show $B_{\mathrm{ex}}^{\Gamma_{\mathrm{AM}}}/T_K$ as a function of $\phi$, where $\Gamma_{\mathrm{AM}}$ denotes the irrep of the altermagnetic form factor; the shaded and unshaded regions denote the orientations $\phi$ with unsplit and split signals, respectively. Given the input, the output is thus obtained as,
\begin{equation}
Z^{\Gamma_{\mathrm{AM}}}(X,Y)=\Theta\!\left(B_{\mathrm{ex}}^{\Gamma_{\mathrm{AM}}}/T_K-1\right),\quad
\Gamma_{\mathrm{AM}}\in\{B_{1g},B_{2g}\},
\end{equation}
where $\Theta$ is the Heaviside step function. The resulting truth tables are shown in Figs.~\ref{fig:logic}(b) and \ref{fig:logic}(c). A $B_{1g}$ altermagnet realizes an XOR gate: either the $x$- or $y$-oriented drive activates the splitting, whereas the simultaneous $x+y$ drive leaves it unsplit. In contrast, a $B_{2g}$ altermagnet realizes an AND gate: the individual drives remain symmetry forbidden, while their simultaneous application activates the splitting. An $A_{2g}$ $g$-wave altermagnet can further realize a phase-parity XNOR gate, as discussed in Sec.~III of the Supplemental Material.

The impurity setup therefore functions as a prototype spectral Kondo logic element in a Floquet-driven altermagnet. For electronic readout, the same idea can be implemented in a Kondo quantum dot coupled to left and right altermagnetic leads. The conductance then follows the same truth tables: an unsplit Kondo resonance gives a high-conductance state approaching $G_{\mathrm{high}}\to2e^2/h$ in the ideal unitary limit \cite{GlazmanRaikh1988KondoTransport,NgLee1988ResonantTunneling,GoldhaberGordon1998SET}, whereas a split resonance gives $G_{\mathrm{low}}\ll G_{\mathrm{high}}$ \cite{Costi2000MagneticField}. This electronic readout provides a possible route toward cascading Kondo-logic elements, once the output controls the drive of the next stage.

\emph{\color{blue}{Conclusion and discussion.--}}
We reveal an altermagnetic Kondo phenomenon---a controllable Kondo splitting tunable by external AC fields. In equilibrium, a local $s$-wave impurity is symmetry-blind to the momentum-dependent spin splitting because nontrivial altermagnetic form factors cancel in the hybridization. We show that an in-plane AC field acts as a Floquet symmetry converter, lifting this cancellation and generating an $A_{1g}$-symmetric spin-dependent channel. Kondo correlations amplify the resulting local spin contrast into a spin-resolved resonance detectable by STM/STS. Remarkably, its angular dependence inherits the altermagnetic form factor, providing a many-body tomography that reveals not only the spin splitting but also its underlying symmetry. More broadly, the altermagnetic Kondo effect substantially expands the spectroscopic power of a local Kondo impurity. Its field-angle-dependent splitting does not merely identify a symmetry class \cite{Balatsky2006Impurity}; it directly reconstructs the crystalline angular pattern of the momentum-dependent magnetic state [Fig.\ref{fig:kondo}(c) and Fig.\ref{fig:logic}(b)(c)]. This establishes a symmetry-resolved form of Kondo tomography and represents a significant advance in Kondo spectroscopy.

The proposed spectral readout is compatible with recent developments in lightwave-driven STM
\cite{Peller2021Waveform,Roelcke2024UltrafastSTS,Cocker2013UltrafastTHzSTM},
where intense near-field AC waveforms are combined with atomic-scale spatial resolution. Such a setup offers a natural route to apply a local Floquet drive to an altermagnetic surface or impurity region while simultaneously detecting the split or unsplit Kondo resonance. This provides a practical route to resolving the symmetry structure of altermagnetic spin splitting.

Finally, the same split/unsplit response directly enables altermagnetic Kondo logic. Here, the logic operation is not imposed by external circuit elements, but encoded by the symmetry-selected visibility of the spin splitting itself. Extensions to higher irreps, multiorbital altermagnets, and elliptically or circularly polarized drives may further enlarge the family of symmetry-controlled Kondo logic operations. Our work thus establishes Floquet-driven altermagnets as a platform where symmetry, Kondo correlations, and logic functionality are intrinsically intertwined.

 \begin{acknowledgments}
Haojie Shen and Xinchen Zhou contributed equally to this work. This work was supported by the National Key R\&D Program of China (Grant No. 2022YFA1403601, 2024YFA1410500), HFNL Self-Deployed Project (Grant No. ZB2602000303), the Scientific Research Innovation Capability Support Project for Young Faculty (Grant No. SRICSPYF-ZY2025164), the National Natural Science Foundation of China (Grant No.12322402, No.12274206), the Quantum Science and Technology-National Science and Technology Major Project (Grant No.2021ZD0302800), the Natural Science Foundation of Jiangsu Province (Grant No.BK20233001), and the Fundamental Research Funds for the Central Universities (Grant No. KG202501).
 \end{acknowledgments}


\begin{thebibliography}{99}
\bibitem{footnote0}
The dominating coupling channel between an impurity and bath electrons is
isotropic $s$-wave for common cases. Although anisotropy-induced higher
partial-wave couplings could exist in realistic materials, they are weak
compared to the $s$-wave channel and can be usually neglected.
\end{thebibliography}
\end{document}


\title{Supplemental Material for ``Altermagnetic Kondo Logic via Floquet Symmetry Conversion''}

\author{Haojie Shen}
\affiliation{National Laboratory of Solid State Microstructures and Department of Physics, Nanjing University, Nanjing 210093, China}

\author{Xinchen Zhou}
\affiliation{National Laboratory of Solid State Microstructures and Department of Physics, Nanjing University, Nanjing 210093, China}

\author{Baigeng Wang}
\affiliation{National Laboratory of Solid State Microstructures and Department of Physics, Nanjing University, Nanjing 210093, China}
\affiliation{Collaborative Innovation Center of Advanced Microstructures, Nanjing University, Nanjing 210093, China}
\affiliation{Jiangsu Physical Science Research Center, Nanjing University, Nanjing 210093, China}

\author{Rui Wang}
\email{rwang89@nju.edu.cn}
\affiliation{National Laboratory of Solid State Microstructures and Department of Physics, Nanjing University, Nanjing 210093, China}
\affiliation{Collaborative Innovation Center of Advanced Microstructures, Nanjing University, Nanjing 210093, China}
\affiliation{Jiangsu Physical Science Research Center, Nanjing University, Nanjing 210093, China}
\affiliation{Hefei National Laboratory, Hefei 230088, People's Republic of China}

\date{July 28, 2026}
\begin{abstract}
We provide supplemental information regarding all technical details on the proofs and derivations of key conclusions presented in the manuscript.  Specific contents include: 1) Floquet engineering of the local hybridization function, 2) Square-lattice realization and NRG benchmark, and 3) $A_{2g}$ phase-parity XNOR gate for $g$-wave altermagnets. 
\end{abstract}
\maketitle
\tableofcontents

\section{Floquet engineering of the local hybridization function}
\label{sec:continuum}

Throughout this supplemental material, the external field is described by the spatially uniform vector potential $\ba(t)$, which generates the periodic electric field via $\mathbf E(t)\propto-\partial_t\ba(t)$.  The field orientation is labeled by the angle $\phi$.
Sec.~\ref{sec:continuum}  derives a low-energy continuum model of altermagnets driven by Floquet field, which is coupled to an Anderson impurity. We will clearly demonstrate analytic selection rules and local-hybridization responses, as concrete examples of the general symmetry framework presented in the main text.  Sec.~\ref{sec:sm_lattice_nrg} focuses on the lattice model and the details of NRG calculations. More generalizations of our framework to $g$-wave altermagnets and the discussion of the corresponding logic gates will also be presented in Sec.~\ref{sec:sm_lattice_nrg}.

\subsection{Minimal altermagnetic Anderson model}

The minimal continuum description of $d$-wave altermagnets with momentum-dependent spin-splitting band structures is given by 
\cite{Smejkal2022PRX,Brekke2023PRB},
\begin{equation}
    H_0
    =
    \sum_{\bk\sigma}
    \varepsilon_{\bk\sigma}^{0}
    c_{\bk\sigma}^{\dagger}c_{\bk\sigma},
    \qquad
    \varepsilon_{\bk\sigma}^{0}
    =
    \xi_{\bk}+\sigma M_{\bk},
    \qquad
    \xi_{\bk}
    =
    \frac{\hbar^2k^2}{2m}-\mu,
    \qquad
    \sigma=\pm1 .
    \label{eq:sm_ham_am}
\end{equation}
Here $\sigma=+1$ and $-1$ denote spin-up and spin-down.  For a continuum $d$-wave altermagnet,
\begin{equation}
    M_{\bk}
    =
    M\frac{k^2}{k_F^2}
    \cos[2(\theta_{\bk}-\theta_d)] ,
    \label{eq:sm_m_polar}
\end{equation}
where $M$ describes the spin-splitting strength.   Equivalently, the form factor can be written as
\begin{equation}
    M_{\bk}
    =
    \frac{M}{k_F^2}
    \left[
        (k_x^2-k_y^2)\cos2\theta_d
        +
        2k_xk_y\sin2\theta_d
    \right].
    \label{eq:sm_m_cart}
\end{equation}
In the spinor basis of spin-up and spin-down electrons, the spin-dependent dispersion is given by the Hamiltonian in the matrix form as,
\begin{equation}
    H_0(\bk)
    =
    \xi_{\bk}\sigma_0+M_{\bk}\sigma_z .
    \label{eq:sm_band}
\end{equation}
Note that the altermagnetic form factor satisfies
\begin{equation}
    M_{R_{\pi/2}\bk}=-M_{\bk}.
    \label{eq:sm_d_wave_sign}
\end{equation}
Thus, the finite spin-splitting at a fixed momentum changes sign under a $\pi/2$ rotation.

We then consider an Anderson impurity described by
\begin{equation}
    H_{\rm imp}
    =
    \sum_{\sigma}
    \epsilon_d d_{\sigma}^{\dagger}d_{\sigma}
    +
    U n_{d\uparrow}n_{d\downarrow},
    \qquad
    n_{d\sigma}=d_{\sigma}^{\dagger}d_{\sigma},
    \label{eq:sm_ham_imp}
\end{equation}
which is  locally coupled to the altermagnetic bath through an isotropic
$s$-wave hybridization $V_0$, namely
\begin{equation}
    H_{\rm hyb}
    =
    V_0
    \sum_{\bk\sigma}
    \left(
        c_{\bk\sigma}^{\dagger}d_{\sigma}
        +
        d_{\sigma}^{\dagger}c_{\bk\sigma}
    \right).
    \label{eq:sm_ham_hyb}
\end{equation}
As in the standard treatment of $s$-wave quantum impurities, here only the local symmetric component of the bath Green's function enters into the self-energy correction of the impurity.

\subsection{Static cancellation in the local hybridization}

The bath enters the impurity action through the hybridization function
\begin{equation}
    \Delta_{\sigma}(z)
    =
    \sum_{\bk}
    \frac{|V_0|^2}{z-\xi_{\bk}-\sigma M_{\bk}}.
    \label{eq:sm_delta_static}
\end{equation}
The spin contrast of the hybridization function is then given by
\begin{equation}
    \delta\Delta(z)
    \equiv
    \Delta_{\uparrow}(z)-\Delta_{\downarrow}(z),
    \label{eq:sm_delta_static_difference_definition}
\end{equation}
which is evaluated as
\begin{align}
    \delta\Delta(z)
    &=
    \sum_{\bk}|V_0|^2
    \left[
        \frac{1}{z-\xi_{\bk}-M_{\bk}}
        -
        \frac{1}{z-\xi_{\bk}+M_{\bk}}
    \right]
    \notag\\
    &=
    \sum_{\bk}|V_0|^2
    \frac{2M_{\bk}}
    {(z-\xi_{\bk})^2-M_{\bk}^2}.
    \label{eq:sm_delta_static_difference}
\end{align}
For a continuum bath,  the sum of momentum can be written into the integral,   
\begin{equation}
    \delta\Delta(z)
    =
    2|V_0|^2
    \int\frac{k\dd k}{(2\pi)^2}
    \int_0^{2\pi}\dd\theta_{\bk}\,
    \frac{M_{\bk}}
    {(z-\xi_k)^2-M_{\bk}^2}.
    \label{eq:sm_static_integral}
\end{equation}
We denote the integrand by $f_k(\theta_{\mathbf{k}})$, i.e.,
\begin{equation}
    f_k(\theta_{\bk})
    =
    \frac{M_{\bk}}
    {(z-\xi_k)^2-M_{\bk}^2}
    \label{eq:sm_fk},
\end{equation}
which clearly satisfies
\begin{equation}
    f_k(\theta_{\bk}+\pi/2)=-f_k(\theta_{\bk}).
    \label{eq:sm_fk_odd}
\end{equation}
Consequently, it is found that,
\begin{equation}\label{eq:sm_angular_cancel1}
    \int_0^{2\pi}\dd\theta_{\bk}\,f_k(\theta_{\bk})=
    \int_0^{\pi/2}\dd\theta_{\bk}
    \sum_{\ell=0}^{3}
    f_k(\theta_{\bk}+\ell\pi/2)=0,
\end{equation}
and hence
\begin{equation}
    \Delta_{\uparrow}(z)=\Delta_{\downarrow}(z).
    \label{eq:sm_static_spin_blind}
\end{equation}
Therefore, the spin contrast hybridization function $\delta\Delta(z)$ vanishes for the undriven or static case, as a result of the symmetry enforced cancellation. 

\subsection{Off-resonant Floquet dressing}

A spatially uniform Floquet drive enters the bath through the Peierls substitution,
\begin{equation}
    \bk\rightarrow\bk+\ba(t).
    \label{eq:sm_peierls}
\end{equation}
We will show in the following that, in the off-resonant high-frequency regime, the Floquet dressing has two relevant consequences. First, the cycle average modifies the dispersion and magnetic form factor of the altermagnets. Second, the virtual processes generated by nonzero Fourier harmonics give subleading corrections to the impurity coupling.
We write the external vector potential in the form,
\begin{equation}
    a_i(t)
    =
    {\rm Re}\,[\tilde a_i\ee^{-\ii\Omega t}],
    \qquad
    i=x,y,
    \label{eq:sm_complex_drive_amplitude}
\end{equation}
where $\langle a_i(t)\rangle_T=0$, and  $T=2\pi/\Omega$ is the drive period. The corresponding single-particle bath Hamiltonian is then cast into,
\begin{equation}
    H(\bk,t)
    =
    \xi_{\bk+\ba(t)}\sigma_0
    +
    M_{\bk+\ba(t)}\sigma_z .
    \label{eq:sm_driven_ham}
\end{equation}
Then expand the Peierls-shifted form factor can be expanded as,
\begin{align}
    \left\langle M_{\bk+\ba(t)}\right\rangle_T
    &=
    M_{\bk}
    +
    \sum_{n=1}^{\infty}
    \frac{1}{n!}
    Q_{i_1\ldots i_n}^{(n)}
    \partial_{k_{i_1}}\cdots\partial_{k_{i_n}}M_{\bk},
    \notag\\
    Q_{i_1\ldots i_n}^{(n)}
    &\equiv
    \left\langle
        a_{i_1}(t)\cdots a_{i_n}(t)
    \right\rangle_T ,
    \label{eq:sm_general_drive_tensor_expansion}
\end{align}
where $Q_i^{(1)}=0$ because of the zero-mean of the external field,  $\langle a_i(t)\rangle_T=0$.  A monochromatic drive also satisfies $a_i(t+T/2)=-a_i(t)$, so all the odd-rank tensors
$Q^{(2m+1)}$ vanish.  The leading order correction is therefore governed by
\begin{equation}
    C_{ij}\equiv  Q_{ij}^{(2)}
    =
    \langle a_i(t)a_j(t)\rangle_T
   .
    \label{eq:sm_q2_equals_c}
\end{equation}
For continuum $d$-wave form factor $M_{\mathbf{k}}$ quadratic in momentum, the series
terminates at this order.  Higher harmonics can instead require higher-rank
drive tensors, as  the fourth-order $g$-wave example shown in Sec.~\ref{sec:sm_gwave}.

Expanding the time-dependent Hamiltonian in Floquet harmonics,
\begin{equation}
    H(\bk,t)=\sum_n H_n(\bk)\ee^{-\ii n\Omega t},
    \qquad
    H_n(\bk)
    =
    \frac{1}{T}
    \int_0^T\dd t\,H(\bk,t)\ee^{\ii n\Omega t},
    \label{eq:sm_fourier}
\end{equation}
the van Vleck expansion gives the static stroboscopic Hamiltonian
\cite{GoldmanDalibard2014,Bukov2015,EckardtAnisimovas2015}
\begin{equation}
    H_{\rm eff}
    =
    H_{n=0}+
    \sum_{n>0}
    \frac{[H_{-n},H_n]}{n\hbar\Omega}
    +
    O(\Omega^{-2}).
    \label{eq:sm_van_vleck}
\end{equation}
Here $ H_{n=0}$ denotes the cycle average of the Hamiltonian, which contains the Peierls-shifted form factor, $  \left\langle M_{\bk+\ba(t)}\right\rangle_T$, and thus the Floquet drive tensor $Q^{(n)}_{i_1,\ldots i_n}$.
Insertion of Eq.~\eqref{eq:sm_complex_drive_amplitude},
\[
    a_i(t)
    =
    \frac{1}{2}
    \left(
        \tilde a_i\ee^{-\ii\Omega t}
        +
        \tilde a_i^*\ee^{\ii\Omega t}
    \right),
\]
into the drive tensor,  we find both oscillating terms with frequency $2\Omega$ and zero-frequency terms.  Only the zero-frequency part survives after the cycle average, leading to
\begin{equation}
    \begin{aligned}
    C_{ij}
    &=
    \langle a_i(t)a_j(t)\rangle_T=
    \frac{1}{4}
    \left(
        \tilde a_i\tilde a_j^*
        +
        \tilde a_i^*\tilde a_j
    \right)
    =
    \frac{1}{2}{\rm Re}\,
    \left(\tilde a_i\tilde a_j^*\right).
    \end{aligned}
    \label{eq:sm_drive_tensor}
\end{equation}
Thus, the Peierls cycle average retains only the real symmetric part of the
drive bilinears.  In comparison, we term the imaginary antisymmetric part,
\begin{equation}
    Y_{ij}
    =
    {\rm Im}\,
    \left(\tilde a_i\tilde a_j^*\right)
    =
    -Y_{ji}
    \label{eq:sm_helicity_tensor}
\end{equation}
the helicity channel, which measures the oriented area of the drive ellipse. The distinction between $C_{ij}$ and $Y_{ij}$ is clear: $C_{ij}$ enters the zero harmonic $H_{n=0}=\langle H(t)\rangle_T$, while $Y_{xy}$ can only enter the Floquet Hamiltonian through the commutator in
Eq.~\eqref{eq:sm_van_vleck}.  To see this more explicitly, we expand $H(\mathbf{k},t)$ to first order of the drive amplitude, i.e.,
\begin{align}
    H(\bk,t)
    &=
    H_0(\bk)+H^{(1)}(\bk,t)+O(a^2),
    \label{eq:sm_linear_hamiltonian_expansion}
\end{align}
where 
\begin{equation*}
    H^{(1)}(\bk,t)=
    a_i(t)\partial_{k_i}H_0(\bk)
    =
    \frac{1}{2}\tilde a_i\partial_{k_i}H_0(\bk)\ee^{-\ii\Omega t}
    +
    \frac{1}{2}\tilde a_i^*\partial_{k_i}H_0(\bk)\ee^{\ii\Omega t}.
\end{equation*}
Here the superscript denotes the first order expansion.  Comparing
Eq.~\eqref{eq:sm_linear_hamiltonian_expansion} with 
Eq.~\eqref{eq:sm_fourier}, we obtain the  $n=\pm1$ harmonics as,
\begin{equation}
    H_1^{(1)}
    =
    \frac{1}{2}\tilde a_i\partial_{k_i}H_0,
    \qquad
    H_{-1}^{(1)}
    =
    \frac{1}{2}\tilde a_i^*\partial_{k_i}H_0,
    \label{eq:sm_first_harmonics}
\end{equation}
and hence
\begin{equation}
    \frac{[H_{-1}^{(1)},H_1^{(1)}]}{\hbar\Omega}
    =
    -\frac{\ii Y_{xy}}{2\hbar\Omega}
    [\partial_{k_x}H_0,\partial_{k_y}H_0].
    \label{eq:sm_helicity_commutator}
\end{equation}
Thus,  the helicity channel  can only survive when  non-commuting derivative
matrices are present.

For the collinear altermagnets driven by Floquet fields considered in our work, all the Fourier components are linear combinations of $\sigma_0$ and $\sigma_z$.  Hence
\begin{equation}
    [H_{-n}(\bk),H_n(\bk)]=0,
    \label{eq:sm_commutator_zero}
\end{equation}
and, more generally, $[H(\bk,t),H(\bk,t')]=0$.  Therefore, one obtains
the stroboscopic Hamiltonian
\begin{equation}
    H_{\rm eff}(\bk)=H_{n=0}(\bk)=\langle H(\bk,t)\rangle_T.
    \label{eq:sm_cycle_average}
\end{equation}
Note that Eq. \eqref{eq:sm_commutator_zero} may not be exactly satisfied when the low-energy band structure of the altermagnets is further complicated by spin-orbit couplings (SOC). Since it is the altermagnetic spin-splitting rather than the SOC-induced splitting that is the key feature underlying altermagnetic materials, we do not consider such an effect as in most previous theoretical studies.

It is known from above that the Floquet Hamiltonian is determined by the real symmetry drive tensor $C_{ij}$ . Notably, it can be further written into a sum of components, i.e.,   $A_{1g}\oplus B_{1g}\oplus B_{2g}$ , corresponding to the irreps of the underlying point group: 
\[
    C
    =
    \frac{Q_{A_{1g}}^{(2)}}{2}
    \begin{pmatrix}
        1 & 0 \\
        0 & 1
    \end{pmatrix}
    +
    \frac{Q_{B_{1g}}^{(2)}}{2}
    \begin{pmatrix}
        1 & 0 \\
        0 & -1
    \end{pmatrix}
    +
    \frac{Q_{B_{2g}}^{(2)}}{2}
    \begin{pmatrix}
        0 & 1 \\
        1 & 0
    \end{pmatrix}.
\]
The three coefficients are given by
\begin{align}
    Q_{A_{1g}}^{(2)}
    &=
    C_{xx}+C_{yy},
    \notag\\
    Q_{B_{1g}}^{(2)}
    &=
    C_{xx}-C_{yy},
    \notag\\
    Q_{B_{2g}}^{(2)}
    &=
    2C_{xy}.
    \label{eq:sm_quadratic_filters}
\end{align}
Note that the trace $Q_{A_{1g}}^{(2)}$ gives the scalar ponderomotive shift,
\begin{equation}
     \left\langle \xi_{\bk+\ba(t)}\right\rangle_T=
    \xi_{\bk}
    +
    \frac{\hbar^2}{2m}
    \left\langle
        2\bk\cdot\ba(t)+a^2(t)
    \right\rangle_T
    =
    \xi_{\bk}
    +
    \frac{\hbar^2}{2m}
    Q_{A_{1g}}^{(2)}
    \equiv
    \xi_{\bk}^{F}.
    \label{eq:sm_xi_average}
\end{equation}

This term can be absorbed into the chemical potential in the continuum model. 

The altermagnetic form factor receives  nontrivial spin-dependent corrections.
The first-order term in Eq. \eqref{eq:sm_general_drive_tensor_expansion} vanishes under the cycle average. For the harmonic continuum $d$-wave form factor, we have $(\partial_{k_x}^2+\partial_{k_y}^2)M_{\bk}=0$, so the trace $Q_{A_{1g}}^{(2)}$ does not generate any uniform spin-dependent corrections. One can straightforwardly obtain
\begin{align}
    \left\langle M_{\bk+\ba(t)}\right\rangle_T
    &=
    M_{\bk}
    +
    \frac{1}{2}C_{ij}\partial_{k_i}\partial_{k_j}M_{\bk}
    =
    M_{\bk}+M_{\Omega},
    \notag\\
    M_{\Omega}
    &=
    \frac{M}{k_F^2}
    \left[
        Q_{B_{1g}}^{(2)}\cos2\theta_d
        +
        Q_{B_{2g}}^{(2)}\sin2\theta_d
    \right].
    \label{eq:sm_m_omega_general}
\end{align}

We now specialize to the linearly polarized drive, which is the main focus of our  main text,
\begin{equation}
    \ba(t)
    =
    a_0\cos\Omega t(\cos\phi,\sin\phi).
    \label{eq:sm_vector_potential}
\end{equation}
Here $a_x^0=a_0\cos\phi$, $a_y^0=a_0\sin\phi$, using which one can obtain
\begin{equation}
    Q_{B_{1g}}^{(2)}
    =
    \frac{a_0^2}{2}\cos2\phi,
    \qquad
    Q_{B_{2g}}^{(2)}
    =
    \frac{a_0^2}{2}\sin2\phi .
    \label{eq:sm_linear_quadrupoles}
\end{equation}
Eq. \eqref{eq:sm_m_omega_general} then gives
\begin{equation}
    M_{\Omega}
    =
    \frac{M}{2}\alpha^2\cos[2(\theta_d-\phi)].
    \label{eq:sm_m_omega_alpha}
\end{equation}
where $\alpha=a_0/k_F$. Thus, it becomes clear that the Floquet linear drives along $x$ and $y$  orientations give opposite responses, while
a diagonal drive is an invariant ``node" for the $B_{1g}$ $d$-wave altermagnets.

For comparison, consider the general elliptic drive, i.e.,
\begin{equation}
    a_x(t)=a_x^0\cos\Omega t,
    \qquad
    a_y(t)=a_y^0\cos(\Omega t+\delta),
    \label{eq:sm_general_elliptic_drive}
\end{equation}
corresponding to $\tilde a_x=a_x^0$ and
$\tilde a_y=a_y^0\ee^{-\ii\delta}$.  The corresponding real symmetric tensor is obtained as
\begin{equation}
    C
    =
    \frac{1}{2}
    \begin{pmatrix}
        (a_x^0)^2 & a_x^0a_y^0\cos\delta \\
        a_x^0a_y^0\cos\delta & (a_y^0)^2
    \end{pmatrix},
    \label{eq:sm_drive_tensor_matrix}
\end{equation}
where
\begin{align}
    Q_{A_{1g}}^{(2)}
    &=
    \frac{(a_x^0)^2+(a_y^0)^2}{2},
    \notag\\
    Q_{B_{1g}}^{(2)}
    &=
    \frac{(a_x^0)^2-(a_y^0)^2}{2},\\
    Q_{B_{2g}}^{(2)}
   & =
    a_x^0a_y^0\cos\delta.
    \label{eq:sm_elliptic_quadrupoles}
\end{align}
The imaginary antisymmetric component is similarly derived to be
\begin{equation}
    Y_{xy}=a_x^0a_y^0\sin\delta.
    \label{eq:sm_helicity_component}
\end{equation}
Therefore, a linear drive with $\delta=0$ has $Y_{xy}=0$ and only exhibits the real tensor $C_{ij}$.  A circular drive, $a_x^0=a_y^0$ and $\delta=\pm\pi/2$, has no real quadrupole but has
nonzero helicity.  An elliptic drive can contain both. The polarization angle $\delta$ thereby provides another tuning knobs that further enrich the Floquet engineered impurity physics in altermagents.

Collecting both the dressed form factor and the dispersion term, the Floquet Hamiltonian of the altermagnetic bath then takes the final form,
\begin{equation}
    H^F
    =
    \sum_{\bk\sigma}
    \left[
        \xi_{\bk}^{F}
        +
        \sigma(M_{\bk}+M_{\Omega})
    \right]
    c_{\bk\sigma,0}^{\dagger}c_{\bk\sigma,0}.
    \label{eq:sm_ham_floquet}
\end{equation}
Here we also mention that the $A_{2g}$ component cannot be derived for the $d$-wave altermagnets--the symmetric tensor $C_{ij}$ contains only $A_{1g}\oplus B_{1g}\oplus B_{2g}$. Its lowest matching filter is fourth order, for example
    $Q_{A_{2g}}^{(4)}=\langle a_xa_y(a_x^2-a_y^2)\rangle_T$, which gives the
linear-drive $\sin4\phi$ selection discussed in Sec.~\ref{sec:sm_gwave}. Such a component can only be non-vanishing for altermagnets with nonzero $g$-wave form factors.

We now further consider the coupling of the driven system to an external impurity. The  driven impurity problem is time dependent, given by
\begin{equation}
    H_{\rm tot}(t)
    =
    H_{\rm bath}(t)+H_{\rm imp}+H_{\rm hyb},
    \label{eq:sm_total_driven_impurity}
\end{equation}
where $H_{\rm bath}(t)$ is the Peierls-dressed bath in
Eq.~\eqref{eq:sm_driven_ham}.  Since the Peierls substitution acts in the hopping terms via
$\exp[\ii\ba(t)\cdot(\mathbf R_i-\mathbf R_j)]$ and  the coupling  in Eq.~\eqref{eq:sm_ham_hyb} is onsite without any hopping terms, the drive field does not affect the impurity coupling $H_{\mathrm{hyb}}$.   Thus, both $H_{\rm imp}$ and $H_{\rm hyb}$ are  time-independent; all nonzero Floquet harmonics of $H_{\rm tot}(t)$ reside in the bath sector.

Following the procedures above, the total time-dependent Hamiltonian is written into, $H_{\rm tot}(t)=\sum_n H_{{\rm tot},n}\ee^{-\ii n\Omega t}$, where
\begin{align}
    H_{{\rm tot},n=0}
    &=
    H_{{\rm bath},n=0}+H_{\rm imp}+H_{\rm hyb},
    \label{eq:sm_total_h0}
    \\
    H_{{\rm tot},n}
    &=
    H_{{\rm bath},n},
    \qquad n\neq0 .
    \label{eq:sm_total_hn}
\end{align}
Here the subscript $n$ labels the Fourier harmonics and
$H_{{\rm bath},n=0}=H^F$ is the Floquet Hamiltonian of the altermagnetic bath derived above.

Since the  bath Hamiltonian is spin diagonal and collinear, each $H_{{\rm bath},n}$ is
diagonal in the momentum and spin space, so that all the bath Fourier components commute with one another.  The $O(\Omega^{-1})$ terms in the high-frequency expansion therefore
vanishes even if the impurity coupling is considered, as we always have
\begin{equation}
    \sum_{n>0}\frac{[H_{{\rm tot},-n},H_{{\rm tot},n}]}{n\hbar\Omega}
    =
    \sum_{n>0}
    \frac{[H_{{\rm bath},-n},H_{{\rm bath},n}]}{n\hbar\Omega}
    =
    0 .
    \label{eq:sm_hyb_no_omega_minus_one}
\end{equation}
Thus, up to second order expansion, the effective Floquet Hamiltonian of the total system  is obtained as,
\begin{equation*}
    H_{\rm eff}
    =
    H_{{\rm tot},n=0}+{\cal O}(\Omega^{-2})
    =
    H^F+H_{\rm imp}+H_{\rm hyb}+{\cal O}(\Omega^{-2}).
\end{equation*}
Note that the impurity coupling has a nonzero contribution, which is of the order of  $O(\Omega^{-2})$. This comes into play through the higher order commutators,    $[H_{{\rm tot},-n},[H_{{\rm tot},n=0},H_{{\rm tot},n}]]$, which includes $[H_{{\rm bath},-n},[H_{\rm hyb},H_{{\rm bath},n}]]$.  Nevertheless, these terms are significantly suppressed and can be safely neglected at the high-frequency regime. Therefore, the local hybridization retains the equilibrium
operator form in Eq.~\eqref{eq:sm_ham_hyb}.  This leads to a Floquet Hamiltonian describing 
a standard Anderson impurity locally coupled  to an AC driven altermagnet.

\subsection{Floquet-induced local hybridization}

Using the above derived total Floquet Hamiltonian, the hybridization function introduced in the main text is obtained as,
\begin{equation}
    \widetilde{\Delta}_{\sigma}(z,\mathbf a)
    =
    \sum_{\bk}
    \frac{|V_0|^2}
    {z-\xi_{\bk}^{F}-\sigma(M_{\bk}+M_{\Omega})}.
    \label{eq:sm_delta_floquet}
\end{equation}
Here $\mathbf a=(a_0,\phi)$ denotes the parameters for the linearly
polarized drive.  The driven spin contrast hybridization is defined  as
\begin{equation}
    \delta\widetilde{\Delta}(z,\mathbf a)
    \equiv
    \widetilde{\Delta}_{\uparrow}(z,\mathbf a)
    -
    \widetilde{\Delta}_{\downarrow}(z,\mathbf a).
    \label{eq:sm_delta_tilde_difference_definition}
\end{equation}
and is derived to be of the form,
\begin{equation}
    \delta\widetilde{\Delta}(z,\mathbf a)
    =
    \sum_{\bk}|V_0|^2
    \frac{2(M_{\bk}+M_{\Omega})}
    {(z-\xi_{\bk}^{F})^2-(M_{\bk}+M_{\Omega})^2}.
    \label{eq:sm_delta_floquet_difference}
\end{equation}
It is clear that, under a $\pi/2$ momentum rotation, the numerator is turned into
$-M_{\bk}+M_{\Omega}$, such that the cancellation as that for the undriven case is now lifted whenever $M_\Omega\neq0$.

The leading dependence on the drive field can be extracted by expanding
Eq.~\eqref{eq:sm_delta_floquet_difference} in $M_{\Omega}$.  Introducing
$X_{\bk}=z-\xi_{\bk}^{F}$, we arrive at
\begin{equation} \label{eq:sm_delta_linear_momega}
    \delta\widetilde{\Delta}(z,\mathbf a)=
    \sum_{\bk}|V_0|^2
    \frac{2(M_{\bk}+M_{\Omega})}
    {X_{\bk}^2-(M_{\bk}+M_{\Omega})^2}
\simeq
    M_{\Omega}
    \sum_{\bk}|V_0|^2
    \frac{2(X_{\bk}^2+M_{\bk}^2)}
    {(X_{\bk}^2-M_{\bk}^2)^2}.
\end{equation}
The zeroth-order term vanishes through Eq.~\eqref{eq:sm_angular_cancel1}.  Combining
Eqs.~\eqref{eq:sm_m_omega_alpha} and \eqref{eq:sm_delta_linear_momega} gives
\begin{equation}
    \delta\widetilde{\Delta}(z,\mathbf a)
    \propto
    a_0^2\cos[2(\theta_d-\phi)].
    \label{eq:sm_delta_selection}
\end{equation}
The corresponding broadening function is
\begin{equation}
    \widetilde{\Gamma}_{\sigma}(\omega,\mathbf a)
    =
    -{\rm Im}\,
    \widetilde{\Delta}_{\sigma}(\omega+\ii0^+,\mathbf a)
    =
    \pi\sum_{\bk}|V_0|^2
    \delta[
        \omega-\xi_{\bk}^{F}
        -
        \sigma(M_{\bk}+M_{\Omega})
    ] .
    \label{eq:sm_gamma_floquet}
\end{equation}
The spin contrast of the broadening function introduced in the main text is
\begin{equation}
    \delta\Gamma(\omega,a_0,\phi)
    =
    -{\rm Im}\,
    \delta\widetilde{\Delta}(\omega+\ii0^+,a_0,\phi)
    =
    \widetilde{\Gamma}_{\uparrow}(\omega,\mathbf a)
    -
    \widetilde{\Gamma}_{\downarrow}(\omega,\mathbf a).
    \label{eq:sm_delta_gamma}
\end{equation}
Note that the real part of $\delta\widetilde{\Delta}$ generates a bath-induced effective exchange field quantified by $B_{\rm ex}$ in the main text. This drives a spin-splitting impurity spectrum,  tunable by the Floquet field.

So far, we have presented the Floquet symmetry conversion mechanism taking a specific $d$-wave altermagnet as an example. It is proved that the resulting impurity spectrum response should follow a $a_0^2$ amplitude scaling, a $\cos[2(\theta_d-\phi)]$ angular dependence, symmetry-enforced ``nodes", and sign reversals under $\pi/2$ rotations.  

\section{Square-lattice realization and NRG benchmark}
\label{sec:sm_lattice_nrg}

In this section, we start from a full square-lattice model description of altermagnets, and present the detailed NRG calculation results of the Kondo resonance of a quantum impurity.   The results are consistent with those obtained from the continuum model discussed in the last section. The two descriptions  share the same symmetry-encoded altermagnetic Kondo effect, as will be shown in the following.

\subsection{\texorpdfstring{$d$}{d}-wave lattice altermagnetic bath}

The square-lattice bath used for the $d$-wave benchmark is
\begin{equation}
    \varepsilon_{\bk\sigma}^{0}
    =
    -2t(\cos k_x+\cos k_y)-\mu
    +
    \sigma M(\cos k_x-\cos k_y).
    \label{eq:sm_lattice_static_dispersion}
\end{equation}
The drive enters into the bath via 
\begin{equation}
    k_x\rightarrow k_x+a_x(t),
    \qquad
    k_y\rightarrow k_y+a_y(t),
    \label{eq:sm_lattice_peierls}
\end{equation}
where
\begin{equation}
    a_x(t)=a_0\cos\Omega t\cos\phi,
    \qquad
    a_y(t)=a_0\cos\Omega t\sin\phi .
    \label{eq:sm_lattice_drive_components}
\end{equation}
Because the Hamiltonian remains diagonal in the spin basis, the leading
high-frequency Floquet Hamiltonian is given by the cycle average.  Using the identity
\begin{equation}
    \left\langle
        \cos[k_i+a_i\cos\Omega t]
    \right\rangle_T
    =
    J_0(a_i)\cos k_i,
    \label{eq:sm_lattice_bessel_average}
\end{equation}
we obtain
\begin{align}
    \varepsilon_{\bk\sigma}^{F}
    &=
    -2t
    \left(
        {\cal J}_x\cos k_x
        +
        {\cal J}_y\cos k_y
    \right)
    -\mu  +
    \sigma M
    \left(
        {\cal J}_x\cos k_x
        -
        {\cal J}_y\cos k_y
    \right),
    \label{eq:sm_lattice_dispersion}
\end{align}
where
\begin{equation}
    {\cal J}_x=J_0(a_0\cos\phi),
    \qquad
    {\cal J}_y=J_0(a_0\sin\phi).
    \label{eq:sm_lattice_bessel}
\end{equation}
The spin-dependent form factor decomposes as
\begin{align}
    M_{\bk}^{F}
    &=
    M({\cal J}_x\cos k_x-{\cal J}_y\cos k_y)=
    M_d^{F}(\cos k_x-\cos k_y)
    +
    M_s^{F}(\cos k_x+\cos k_y),
    \label{eq:sm_lattice_ds}
\end{align}
where
\begin{equation}
    M_d^{F}
    =
    \frac{M}{2}({\cal J}_x+{\cal J}_y),
    \qquad
    M_s^{F}
    =
    \frac{M}{2}({\cal J}_x-{\cal J}_y).
    \label{eq:sm_lattice_ms_md}
\end{equation}
The first term retains the original $B_{1g}$ altermagnetic form factor, whereas
the second is an $A_{1g}$ spin-dependent component visible to the local
$s$-wave impurity projection.  At weak drive, the long-wavelength expansion gives 
\begin{align}
    {\cal J}_x-{\cal J}_y
    &=
    -\frac{a_0^2}{4}\cos2\phi
    +O(a_0^4),
    \notag\\
    M_s^F(\cos k_x+\cos k_y)
    &=
    -\frac{Ma_0^2}{4}\cos2\phi
    +O(a_0^2k^2,a_0^4).
    \label{eq:sm_lattice_uniform_weak_drive}
\end{align}
Thus the lattice model reproduces the  $\cos2\phi$-dependence derived from the continuum model, i.e., Eq.~\eqref{eq:sm_m_omega_alpha}.

After considering the impurity coupling to the bath, the bath enters into the impurity physics via the hybridization function, 
\begin{equation}
    \Delta_{\sigma}^{\rm lat}(z)
    =
    \frac{|V_0|^2}{N_k^2}
    \sum_{\bk}
    \frac{1}{z-\varepsilon_{\bk\sigma}^{F}} .
    \label{eq:sm_lattice_delta}
\end{equation}
The sum over momentum is done on a momentum grid in the Brillouin zone with $N_k$ sites, and the superscript ``lat" denotes the lattice model. Evaluating the retarded hybridization function with $z=\omega+\ii\eta$, one then obtains
\begin{equation}
    \Gamma_{\sigma}^{\rm lat}(\omega)
    =
    \frac{|V_0|^2}{N_k^2}
    \sum_{\bk}
    \frac{\eta}
    {(\omega-\varepsilon_{\bk\sigma}^{F})^2+\eta^2}.
    \label{eq:sm_lattice_gamma}
\end{equation}
In the standard convention, the Wilson-chain in the NRG calculation is determined from 
\begin{equation}
    \rho_{{\rm hyb},\sigma}(\omega)
    =
    \frac{\Gamma_{\sigma}^{\rm lat}(\omega)}{\pi}
    \label{eq:sm_lattice_rho}
\end{equation}
via an energy discretization and a systematic mapping procedure.  The representative parameters used for the Wilson-chain benchmark are listed in the caption of Fig.~\ref{fig:sm_mapping_recovery}.  We also mention that, for the NRG calculations shown below, we  retain the full Bessel function form in Eq.~\eqref{eq:sm_lattice_bessel} without resorting to the long-wavelength expansion. 

\subsection{Numerical details}
\label{sec:sm_mapping}

We study the mapped Anderson impurity model by FDM NRG.  For each set of drive
parameters $\mathbf a=(a_0,\phi)$, the spin-resolved local hybridization
function obtained after the lattice momentum sum is logarithmically
discretized on its asymmetric support $[\omega_{\min},\omega_{\max}]$ and
mapped onto a semi-infinite spin-diagonal Wilson chain with exponentially
decaying hoppings.  The impurity
couples to the first chain site, and iterative diagonalization with
high-energy-state truncation accesses progressively lower energy scales
\cite{Wilson1975RMP,KrishnaMurthy1980PRB1,KrishnaMurthy1980PRB2,Bulla2008RMP}.
Impurity spectra are computed with the full-density-matrix formulation
\cite{Hofstetter2000PRL,AndersSchiller2005PRL,Peters2006PRB,
Weichselbaum2007PRL,Zitko2009PRB}.  We define
$D_{\rm raw}=\max(|\omega_{\min}|,|\omega_{\max}|)$, store the exported
Wilson-chain coefficients as $e_{n\sigma}/D_{\rm raw}$ and
$t_{n\sigma}/D_{\rm raw}$, and set the NRG energy unit to $D=D_{\rm raw}$.
The dimensionless frequency is
\begin{equation}
    x
    \equiv
    \frac{\omega_{\rm raw}}{D_{\rm raw}}
    =
    \frac{\omega}{D}.
    \label{eq:sm_dimensionless_frequency}
\end{equation}

\begin{figure}[!htbp]
    \centering
    \includegraphics[width=0.68\textwidth]{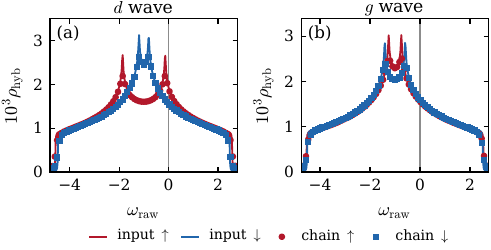}
    \caption{
        Wilson-chain mapping benchmark at $a_0=1$ and $\phi=\pi/8$.
        Solid lines show the raw hybridization functions
        $\Gamma_{\sigma}^{\rm raw}/\pi$, and symbols show their Wilson-chain
        reconstructions for (a) $d$-wave and (b) $g$-wave altermagnetic baths.
    }
    \label{fig:sm_mapping_recovery}
\end{figure}

Figure~\ref{fig:sm_mapping_recovery} shows the first benchmark. It compares the raw hybridization functions
with their Wilson-chain reconstructions for representative $d$- and $g$-wave
baths, directly testing the mapping used in the FDM NRG calculation.

The FDM NRG parameters are
\begin{equation}
    U/D=0.015,\qquad
    \epsilon_d/D=-0.010,\qquad
    b=0.40,\qquad
    \Lambda=1.4,\qquad
    N_z=48.
    \label{eq:sm_nrg_parameters}
\end{equation}
The $z$-trick averages over $N_z$ shifted logarithmic meshes,
$z_j=(j-1/2)/N_z$ with $j=1,\ldots,N_z$, to reduce discretization artifacts. The finite-size level spacing of a chain ending at site $n$ is of order
$\omega_n\propto\Lambda^{-n/2}$, where $\Lambda>1$ is the logarithmic
discretization parameter, and $b$ is the broadening factor as will be introduced below.  We use the first $N_{\rm site}=120$ sites of each
$n_{\rm site}=130$ exported chain and retain up to $1600$ states at each
truncation step.  The impurity lies in the asymmetric local-moment regime,
$\epsilon_d<0<\epsilon_d+U$, with representative zero-energy linewidths
$\gamma_\sigma(0)\sim10^{-3}$.  The zero-temperature impurity spectrum is
broadened as
\begin{equation}
    \delta(\omega-E)
    \longrightarrow
    \frac{e^{-b^2/4}}{b|E|\sqrt{\pi}}
    \exp\!\left[-\left(\frac{\ln|\omega/E|}{b}\right)^2\right],
    \qquad E\ne0,\quad \omega E>0.
    \label{eq:sm_log_gaussian_broadening}
\end{equation}
where $b=0.40$.  The broadening is applied separately to the positive- and
negative-frequency branches and vanishes for $\omega E<0$.  
We denote the unnormalized impurity spectral function in raw units by
$A_{\sigma}^{\rm raw}(\omega_{\rm raw})$ and plot the dimensionless
linewidth-normalized spectrum
\begin{equation}
    \calA_{\sigma}(x)
    =
    \pi\gamma_{\sigma}(x)D_{\rm raw}
    A_{\sigma}^{\rm raw}(D_{\rm raw}x),
    \qquad
    \gamma_{\sigma}(x)
    =
    \frac{\Gamma_{\sigma}^{\rm raw}(D_{\rm raw}x)}{D_{\rm raw}}.
    \label{eq:sm_scaled_spectral_function}
\end{equation}
The plotted frequency is $\omega/D=x$.

\begin{figure}[!htbp]
    \centering
    \includegraphics[width=0.64\textwidth]{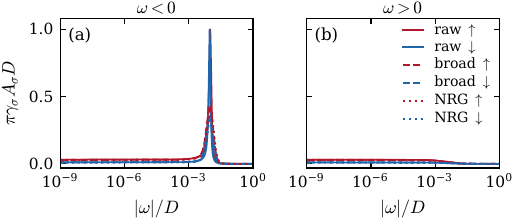}
    \caption{
       The noninteracting ($U=0$) benchmark for the  $d$-wave altermagnetic bath at $a_0=1$,
        $\phi=0$, and $\epsilon_d/D=-0.010$.  The exact spectrum, the same
        spectrum after logarithmic Gaussian broadening with $b=0.40$, and the
        $N_z=48$ FDM NRG result agree with one another throughout the region above the  energy resolution window,
        $\omega_{\rm res}/D=\Lambda^{-N_{\rm site}/2}$.
    }
    \label{fig:sm_u0_benchmark}
\end{figure}

The second benchmark is the noninteracting resonant-level limit.  For $U=0$
and the same driven projected bath,
\begin{equation}
    G_{\sigma}^{(0)}(\omega)
    =
    \frac{1}
    {\omega-\epsilon_d-
    \widetilde{\Delta}_{\sigma}(\omega+\ii0^+,\mathbf a)},
    \qquad
    A_{\sigma}^{(0)}(\omega)
    =
    -\frac{1}{\pi}{\rm Im}\,G_{\sigma}^{(0)}(\omega).
    \label{eq:sm_u0_green}
\end{equation}
Figure~\ref{fig:sm_u0_benchmark} compares the exact spectrum, the same spectrum
broadened with the logarithmic Gaussian function in
Eq.~\eqref{eq:sm_log_gaussian_broadening}, and the FDM NRG result obtained from
the Wilson chain, testing the mapping, normalization, and broadening without
many-body uncertainty.

\begin{figure}[!htbp]
    \centering
    \includegraphics[width=0.80\textwidth]{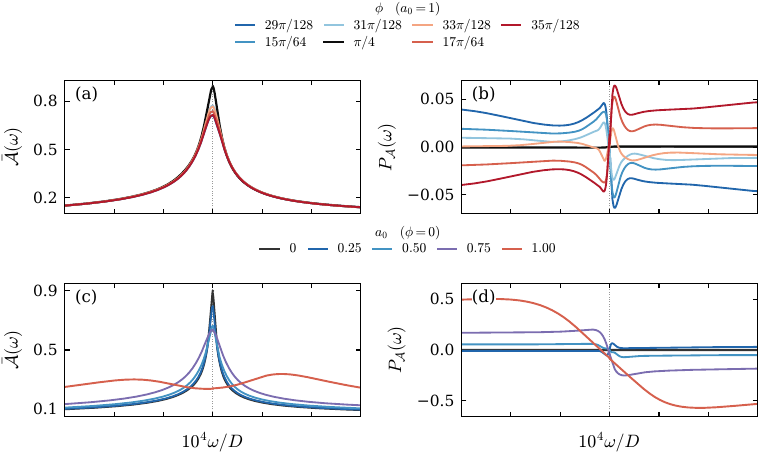}
    \caption{
        Finite-$U$ FDM NRG spectra for $d$-wave altermagnets.  (a) (b) are the spin averaged and spin contrast spectra for angles near the orientation node $\pi/4$  at $a_0=1$, respectively. The spin averaged spectra remains unsplit as $\phi$ is close to $\pi/4$, whereas the spin contrast spectra show spin resolved signals.  (c) (d) are the same spectra but for different  amplitudes  $a_0$ at
        $\phi=0$.  
    }
    \label{fig:sm_nearnull_phi_dense}
\end{figure}

\begin{figure}[!htbp]
    \centering
    \includegraphics[width=0.64\textwidth]{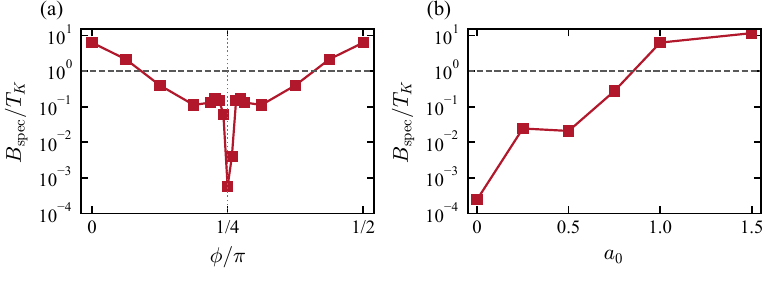}
    \caption{
  The ratio between the maxima distance and the Kondo temperature, $B_{\rm spec}/T_K$, as defined in Eq. \eqref{eq:sm_bspec}.  (a) $B_{\rm spec}/T_K$ as a function of the field orientation $\phi$ at
        $a_0=1$. (b) The amplitude dependence at $\phi=0$.  The dashed lines mark
        $B_{\rm spec}=T_K$, where $T_K$ is determined from the impurity spectrum for $(a_0,\phi)=(1,\pi/4)$.
    }
    \label{fig:sm_bspec_check}
\end{figure}

Figure~\ref{fig:sm_nearnull_phi_dense} shows the finite-$U$  impurity
spectra for $d$-wave altermagnets under angle and drive-amplitude scans. For spectra with two resolved spin-dependent maxima, we also calculate
\begin{equation}
    B_{\rm spec}
    =
    \frac{1}{2}
    \left|
    \omega_{\uparrow}^{\rm pk}
    -
    \omega_{\downarrow}^{\rm pk}
    \right| .
    \label{eq:sm_bspec}
\end{equation}
The ratio between this quantity and $T_K$ provides another transparent diagnosis of whether the spectrum splits or not.  As shown in Fig.~\ref{fig:sm_bspec_check} , $B_{\mathrm{spec}}/T_K$ shows a consistent angle and drive-amplitude dependence as that in Fig.~3(d),(e) of the main text: both are suppressed at the symmetry nodes and become finite in the active channels.

\subsection{\texorpdfstring{$g$}{g}-wave altermagnetic extension}
\label{sec:sm_gwave}

The same impurity protocol can be used as a symmetry filter for higher
altermagnetic harmonics.  A square-lattice representative of a $g$-wave
altermagnet is
\begin{equation}
    \varepsilon_0(\bk)
    =
    -2t(\cos k_x+\cos k_y)-\mu,
    \qquad
    M_g(\bk)
    =
    M_g\sin k_x\sin k_y(\cos k_x-\cos k_y),
    \label{eq:sm_g_lattice_model}
\end{equation}
with energy dispersion,
\begin{equation}
    \varepsilon_{\bk\sigma}^{g,0}
    =
    \varepsilon_0(\bk)+\sigma M_g(\bk),
    \qquad
    \sigma=\pm1.
    \label{eq:sm_g_lattice_band}
\end{equation}
Around the $\Gamma$ point of the Brillouin zone, the form factor is reduced to,
\begin{equation}
    M_g(\bk)
    =
    \frac{M_g}{2}k_xk_y(k_y^2-k_x^2)
    +
    O(k^6)
    \propto
    k^4\sin4\theta_{\bk},
    \label{eq:sm_g_low_energy}
\end{equation}
which is up to an overall sign convention.

Following the same procedure above,  a linearly polarized off-resonant drive leads to the following Floquet Hamiltonian, 
\begin{equation}
    \varepsilon_{\bk\sigma}^{g,F}
    =
    -2t[J_0(a_x)\cos k_x+J_0(a_y)\cos k_y]-\mu
    +
    \sigma M_g\eta_g^F(\bk;\phi),
    \label{eq:sm_g_effective_band}
\end{equation}
where
\begin{align}
    \eta_g^F(\bk;\phi)
    =
    \frac{1}{4}
    \big[
    &J_0(2a_x-a_y)\cos(2k_x-k_y)
    -
    J_0(2a_x+a_y)\cos(2k_x+k_y)
    \notag\\
    &-
    J_0(a_x-2a_y)\cos(k_x-2k_y)
    +
    J_0(a_x+2a_y)\cos(k_x+2k_y)
    \big].
    \label{eq:sm_g_exact_form}
\end{align}
is  the  cycle-averaged $g$-wave form factor, in which we have used 
\begin{equation}
    \left\langle
    \cos[
        m k_x+n k_y+(m a_x+n a_y)\cos\Omega t
    ]\right\rangle_T
    =
    J_0(m a_x+n a_y)\cos(mk_x+n k_y).
    \label{eq:sm_g_general_bessel}
\end{equation}

Under the long-wavelength expansion with a weak field  $a_0$, it is clear that the constant and the $O(a_0^2)$ term cancel in Eq.~\eqref{eq:sm_g_exact_form} , while the fourth-order term survives, leading to
\begin{align}
    \eta_g^F(\mathbf 0;\phi)
    &=
    \frac{3}{16}a_xa_y(a_y^2-a_x^2)
    +
    O(a_0^6)=
    -\frac{3a_0^4}{64}\sin4\phi
    +
    O(a_0^6).
    \label{eq:sm_g_uniform}
\end{align}
Thus, one knows that the angles $\phi=0,\pi/4,\pi/2$ act as the node orientations exhibiting vanishing spin splitting Kondo features,  while angles, such as $\phi=\pi/8$ and $3\pi/8$, related by a $\pi/4$ rotation have the opposite signs.

\begin{figure}[!htbp]
    \centering
    \includegraphics[width=0.64\textwidth]{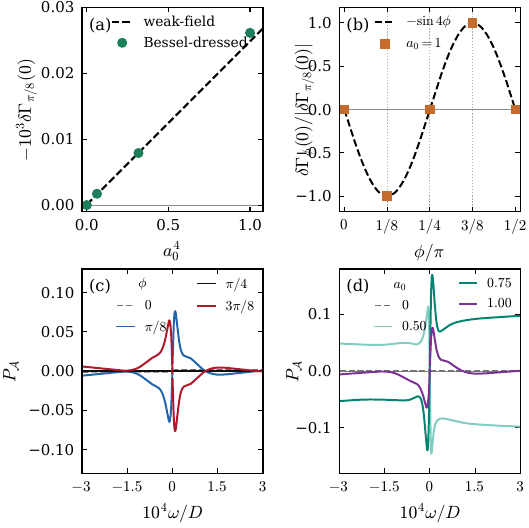}
    \caption{
       NRG results and benchmarks for the magnetic impurity problem in Floquet-driven $g$-wave altermagnets.  (a) shows the spin-splitting broadening function
$\delta\Gamma$ as a function of field amplitude $a_0$ with a fixed orientation $\phi=\pi/8$. A linear growth with  $a_0^4$ is clearly observed. (b) The same as (a) as a function of $\phi$ with a fixed $a_0=1$.  (c) shows the  spin contrast  FDM-NRG spectra $P_{\mathcal{A}}(\omega)$ for different $\phi$ with fixed $a_0=1$, while (d) displays the same as (c) for different $a_0$ with fixed $\phi=\pi/8$.}
    \label{fig:sm_g_selection}
\end{figure}

Similar to the $d$-wave case presented above, the NRG input for the $g$-wave bath  is
\begin{equation}
    \widetilde{\Delta}_{\sigma}^{g}(z,\mathbf a)
    =
    \frac{|V_0|^2}{N_k^2}
    \sum_{\bk}
    \frac{1}{z-\varepsilon_{\bk\sigma}^{g,F}},
    \qquad
    \rho_{{\rm hyb},\sigma}^{g,F}(\omega)
    =
    \frac{\widetilde{\Gamma}_{\sigma}^{g}(\omega,\mathbf a)}{\pi}.
    \label{eq:sm_g_nrg_delta}
\end{equation}
The representative model  parameters used are listed in the caption of
Fig.~\ref{fig:sm_mapping_recovery}, and the NRG parameters are given by
Eq.~\eqref{eq:sm_nrg_parameters}. The final results for the $g$-wave altermagnets are shown in Figure~\ref{fig:sm_g_selection}. They verify the $a^4_0$ scaling behavior, the  angular nodes and the sign reversal feature in Eq. \eqref{eq:sm_g_uniform}, as well as the  spectral response driven by the Kondo correlation.

\subsection[\texorpdfstring{$A_{2g}$}{A2g} XNOR gate]{\texorpdfstring{$A_{2g}$}{A2g} phase-parity XNOR gate for $g$-wave altermagnets}
\label{sec:sm_a2g_xnor}

A phase-parity XNOR gate can  be achieved in an $A_{2g}$ $g$-wave
altermagnet.  Instead of encoding the two inputs in whether the two field
components are present or not, we now encode them in the relative signs of the two AC
components, i.e.,
\begin{equation}
    a_x(t)=s_X a_x^0\cos\Omega t,\qquad
    a_y(t)=s_Y a_y^0\cos\Omega t,
    \qquad
    s_X=(-1)^X,\quad s_Y=(-1)^Y .
    \label{eq:sm_a2g_phase_encoding}
\end{equation}
The $A_{2g}$ drive tensor is then given by
\begin{equation}
    Q_{A_{2g}}^{(4)}
    \equiv
    \langle a_xa_y(a_x^2-a_y^2)\rangle_T
    \propto
    s_Xs_Y a_x^0a_y^0
    \left[(a_x^0)^2-(a_y^0)^2\right].
    \label{eq:sm_a2g_phase_filter}
\end{equation}
For fixed unequal amplitudes and under the convention
$a_x^0a_y^0[(a_x^0)^2-(a_y^0)^2]>0$, the sign of the filter is therefore set by
the phase parity, $s_Xs_Y$.  The superscript on the binary readout labels the
threshold convention, i.e., $Z_{A_{2g}}^{(+)}=1$ for $Q_{A_{2g}}^{(4)}>0$, while
$Z_{A_{2g}}^{(-)}=1$ for $Q_{A_{2g}}^{(4)}<0$.  This then
clearly leads to the truth table in Table~\ref{tab:sm_a2g_xnor}, realizing an XNOR Kondo logic. 

\begin{table}[!htbp]
    \caption{
        The $A_{2g}$ phase-parity encoding and the XNOR Kondo logic gate.  The superscript on
        $Z_{A_{2g}}^{(\pm)}$ indicates which sign of the $A_{2g}$ drive tensor
        is assigned to the output.
    }
    \label{tab:sm_a2g_xnor}
    \centering
    \begin{ruledtabular}
    \begin{tabular}{ccccc}
        $X,Y$ & $s_X,s_Y$ & $\operatorname{sgn}Q_{A_{2g}}^{(4)}$ &
        $Z_{A_{2g}}^{(+)}$ & $Z_{A_{2g}}^{(-)}$ \\
        \hline
        $0,0$ & $+,+$ & $+$ & $1$ & $0$ \\
        $0,1$ & $+,-$ & $-$ & $0$ & $1$ \\
        $1,0$ & $-,+$ & $-$ & $0$ & $1$ \\
        $1,1$ & $-,-$ & $+$ & $1$ & $0$
    \end{tabular}
    \end{ruledtabular}
\end{table}

\clearpage
{\centering\bfseries REFERENCES\par}
\vspace{1.5ex}
\bibliography{supplement_refs}